# Studying cosmic ray sources using intergalactic electromagnetic cascades

A.V. Uryson

P.N. Lebedev Physical Institute of the RAS, Moscow

**Abstract.** In this paper intergalactic electromagnetic cascades are used as a probe of cosmic ray sources. This is achieved as follows. In extragalactic space cosmic rays initiate electromagnetic cascades, in which gamma-ray and neutrino emission arises. We used the joint analysis of cosmic ray data, along with extragalactic gamma-ray and neutrino emissions, to study particle acceleration in the vicinity of supermassive black holes. Particle injection spectra depend on processes of particle acceleration, and here we discuss models with various injection spectra. The computations of the propagation of cosmic rays in space were performed using the publicly available TransportCR code. It was found that a new subclass of sources might exist that does not contribute to the particle flux on Earth, instead to gamma-ray and neutrino emissions arising in electromagnetic cascades. In addition, the upper limit of the relative number of 'exotic' supermassive black holes surrounded by a superstrong magnetic field is derived.

**Keywords:** cosmic rays; electromagnetic cascades; active galactic nuclei; supermassive black hole

### 1. Introduction

Cosmic rays (CR) at ultra-high energies (UHE) are evidently of extragalactic origin, their sources being point-like objects. These objects seem to be active galactic nuclei (AGN), where CRs are accelerated in processes in the vicinity of supermassive black holes (SMBH), which are located in the central parts of AGN. To study UHE CR sources, CR data are used: particle arrival directions, CR energy spectra, and elemental composition. CR arrival directions cannot be used as pointers to sources, due to particle deflection in intergalactic magnetic fields, whose structure is insufficiently studied. Two other parameters change during particle propagation in intergalactic space due to CR interaction with the cosmic microwave background (CMB), extragalactic background light (EBL), and radio background. This results in the GZK-effect [1,2], which is used to study space distribution of CR sources and source models. Another indication of CR interaction with background emissions is intergalactic electromagnetic cascades [3,4]; so now gamma-ray and neutrino emissions produced in cascades are applied to source investigation through the data of the Fermi Large Area Telescope (LAT) onboard the gamma-ray space observatory Fermi, the neutrino observatory IceCube, and the giant ground-based array Pierre Auger Observatory (PAO). The joint analysis of CR data, along with gamma-ray and neutrino emissions, has been performed in a number of articles studying UHE CR source models (see e.g., [5–8]).

In this paper we apply a similar analysis to study CR acceleration in the vicinity of SMBH. CR injection spectra depend on processes in which particles are accelerated, and here we discuss models with various particle spectra. We analyze two CR source models previously presented in [8, 9], which are denoted below by I and II. These models are based on the ideas and results of [10,11].

In the paper [10], electric fields in accretion disks around SMBH are analyzed, and particle acceleration in these fields is discussed. Following the results of [10], we propose model I with hard CR injection spectra, injection spectra being $\propto E^{-a}$, $a \leq 2.2$.

In the paper [11], it is supposed that a superstrong magnetic field of ~$10^{11}$ G possibly exists around SMBH. This idea is argued by some observational data in [12]. The magnetic field induces an electric field, which is not compensated for in some areas of the SMBH magnetosphere and thus accelerates particles. Accordingly, we propose model II, where CR injection spectra are possibly monoenergetic. In this model the contribution to the particle flux is also negligible, as in model I.

It appears that a new subclass of sources possibly exists, showing themselves in gamma-ray and neutrino emissions that CR generate in electromagnetic cascades in intergalactic space. These sources provide no noticeable contribution to the particle flux on Earth. Source characteristics can be studied by measuring the extragalactic diffuse gamma-ray and neutrino background. Analyzing model II using CR and Fermi LAT data, the relative number of SMBHs with superstrong magnetic fields is obtained. The computations of particle propagation in space were performed using the publicly available TransportCR code [13].

## 2. Model I

Currently, AGNs are considered as UHE CR sources. In AGNs, energy-consuming processes are powered by the SMBH action in the galactic center. A number of models have been suggested to describe cosmic ray acceleration in the vicinity of SMBH. In this section we analyze the model [10], in which particles are accelerated by electric fields in the accretion disc. In [10] it is supposed that due to the structure of the field, particles can be accelerated to UHE, the maximal particle energy $E_{max}$ depending linearly on SMBH mass $M$. For several values of SMBH mass, such as $M = 2.5 \cdot 10^6 M_\odot$, $10^7 M_\odot$, $10^8 M_\odot$, where $M_\odot$ is the solar mass, the maximum particle energies are $E_{max} = 10^{20}$, $4 \cdot 10^{20}$, $4 \cdot 10^{21}$ eV, respectively [10]. The local SMBH function being given in [14], we derive that SMBHs with the masses above are distributed with the ratio $2.5 \times 10^6 M_\odot : 10^7 M_\odot : 10^8 M_\odot = 0.313 : 0.432 : 0.254$. In this paper we do not account for CRs from SMBHs with higher and lower masses, as the fraction of these SMBHs is smaller than those above [14].

We consider UHE CR sources with red shifts up to $z = 5.5$, thus it is necessary to account for the evolution of CR sources. The evolution of SMBHs is not yet clear, and we use the evolution scenario of one of the classes of powerful AGNs: Blue Lacertae objects (BL Lac [15], see also [13]).

Next we assume that injection spectra in sources are power-law, $\propto E^{-\alpha}$, with the values of the spectral index $a$=2.2, 1.8, 1, 0.5, 0. Spectra with these indices are harder than those used e.g., in [5], where the main type of possible UHE CR sources is considered. We suppose that the acceleration mechanism suggested in [10] can generate harder spectra. The index value 0 corresponds to the production of an equal number of particles at any UHE.

We assume that UHE CRs consist of protons, and compute only proton propagation. UHE CRs escaping from an acceleration region lose energy in synchrotron and curvature radiation in magnetic fields in the SMBH vicinity. In this paper, these energy losses are assumed to be insignificant. This assumption is based on the case of energy losses of escaping CRs discussed in [16]: approximately 0.4 % of UHE CRs fly away, losing an insignificant part of their energy.

In intergalactic space, UHE CRs interact with CMB, EBL, and radio emissions, for CRs at energies higher than $4 \cdot 10^{19}$ eV the main reactions being $p + \gamma_{background} \rightarrow p + \pi^0$, $p + \gamma_{background} \rightarrow n + \pi^+$. Pions decay $\pi^0 \rightarrow \gamma + \gamma$, $\pi^+ \rightarrow \mu^+ + \nu_\mu$ and give rise to gamma quanta and muons, and muons decay $\mu^+ \rightarrow e^+ + \nu_e + \bar{\nu}_\mu$ giving rise to positrons and neutrinos. CRs at lower energies, of about $10^{18}$ eV, interact with the background via reaction $p + \gamma_{background} \rightarrow p + e^+ + e^-$.

Gamma-quanta and positrons generate electromagnetic cascades interacting with CMB and EBL: $\gamma + \gamma_b \rightarrow e^+ + e^-$ (pair production) and $e + \gamma_b \rightarrow e' + \gamma'$ (inverse Compton effect).

Components of the extragalactic background emission - CMB, EBL, and radio emission, are considered as follows. The CMB has a Planck energy distribution with the mean value $\varepsilon_r = 6.7 \cdot 10^{-4}$ eV, the mean photon density being $n_r = 400$ cm$^{-3}$. The characteristics of EBL are taken from [17]. The background radio emission is described by the model of the luminosity evolution for radio galaxies [18].

In intergalactic space, cascade electrons generate synchrotron emission in magnetic fields and thus lose energy. Synchrotron energy losses are insignificant in fields of $10^{-9}$ G and lower [19]. The intergalactic magnetic field has been insufficiently studied, nevertheless it seems to be nonuniform ([20–22] and ref. in [22]). We suppose that, seemingly, a small part of intergalactic space is occupied by fields higher than $10^{-9}$ G, and they do not disturb cascade development.

### 3. Model II

Model II differs from model I in the point concerning CR sources. In model II we analyze the SMBH hypothesis [11], also discussed in [23], in which charged particles are accelerated in the electric field that is induced near a SMBH by a superstrong magnetic field surrounding it. Particles in [11,23] are accelerated up to $10^{21}$ eV. Based on the acceleration mechanism [11], we assume in model II the CR injection spectrum to be monoenergetic, with the energy E = $10^{21}$ eV.

The cosmic evolution of the CR sources suggested in [11] is associated with the evolution of the states of SMBH (see, e.g., [11]). This evolution being unclear, we consider two possible AGN evolution scenarios: as in BL Lac objects [13,15] or in radio galaxies [24]. In the latter case, we take into account the evolution of only the density of objects, as the luminosity-redshift correlation for radio galaxies is unclear [25].

Other points of model II are the same as in model I.

### 4. Results of model I

The calculated UHE CR energy spectra along with the two spectra measured are shown in Figure 1. The measured spectra were obtained by giant ground-based arrays, the PAO [24] and the Telescope Array (TA) [26]. The calculated spectra are normalized to the spectrum obtained by the PAO at the energy of $10^{19.5}$ eV ($3.16 \cdot 10^{19}$ eV). Each of the calculated spectra are lower than the spectra measured, the difference being of several orders of magnitude (except the point of normalization and the point of about $10^{19.45}$ eV). In addition, the model and measured spectra differ strongly in shape. In model I CRs at energies about $4 \cdot 10^{21}$ eV fall on Earth, but their flux is too low to be detected.

Spectra of the cascade gamma ray emissions are discussed in [27, 28]. In this paper we analyze the intensity of the cascade emission $I_{\text{cascade }\gamma}$ to compare it with Fermi LAT data. Fermi LAT measures the extragalactic isotropic diffuse gamma-ray background (IGRB). This is constituted by several components, two of which are: the cascade emission $I_{\text{cascade }\gamma}$ and the intensity of individual unresolved gamma-ray sources $I_{\text{unresolved blazars}}$. As $I_{\text{cascade }\gamma} + I_{\text{unresolved blazars}} <$ IGRB, the intensity of cascade gamma-ray emission $I_{\text{cascade }\gamma}$ should satisfy the condition:

$$I_{\text{cascade }\gamma} < \text{IGRB} - I_{\text{unresolved blazars}}, \qquad (1)$$

The contribution $P$ of unresolved blazars to IGRB is estimated theoretically [29]: at energies above 50 GeV it is

$$P = 86\,(-14, +16)\%. \tag{2}$$

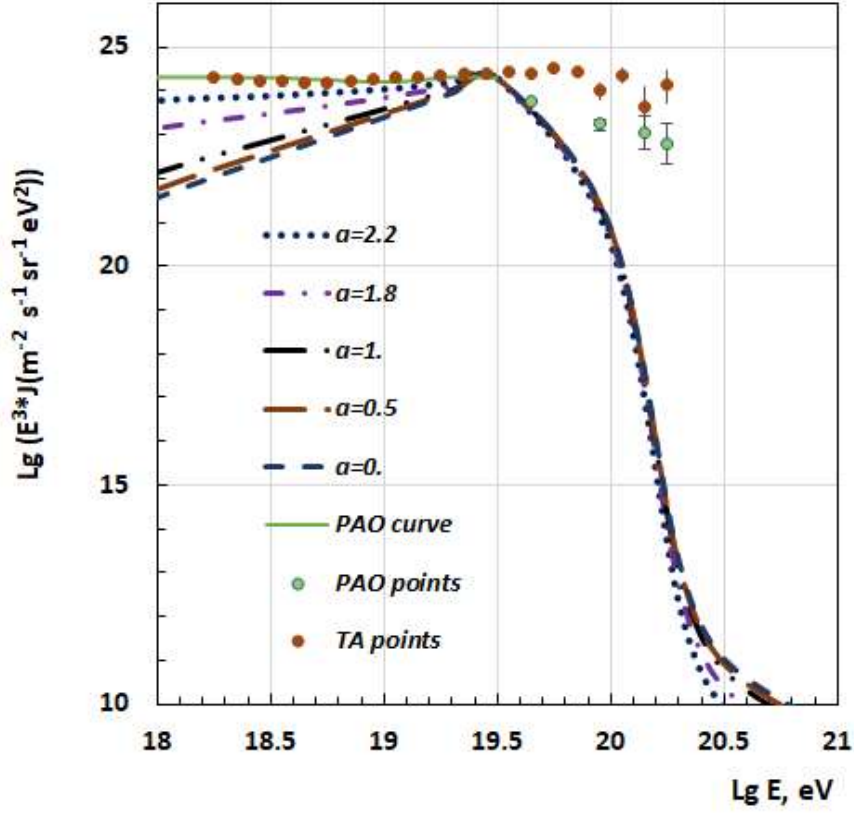

**Figure 1.** Model UHE CR spectra on Earth calculated for injection spectra with various spectral indices a (see the legend), along with the spectra measured on the PAO [24] and the TA [26]. Model spectra are normalized to the PAO data [24] at the energy of $10^{19.5}$ eV.

Therefore, we use model integral intensity $I_{\text{cascade }\gamma}$ in the range $E > 50$ GeV to compare it with Fermi LAT data.

For spectral index values above the cascade integral intensity is:

$$I_{\text{cascade }\gamma}\,(E>50\text{ GeV}) \approx (1.1\text{-}1.6) \times 10^{-10}\,(\text{cm}^{-2}\,\text{s}^{-1}\,\text{sr}^{-1}). \tag{3}$$

IGRB measured by Fermi LAT is [30]:

$$\text{IGRB}\,(E>50\text{ GeV}) = 1.325 \cdot 10^{-9}\,(\text{cm}^{-2}\,\text{s}^{-1}\,\text{sr}^{-1}). \tag{4}$$

Subtracting from the IGRB the unresolved source contribution of 86%, we obtain:

$$\text{IGRB}_{\text{ without unresolved sources}}(E>50\text{ GeV}) = 1.855 \times 10^{-10}\,(\text{cm}^{-2}\,\text{s}^{-1}\,\text{sr}^{-1}). \tag{5}$$

From (3) and (5), the model cascade gamma ray emission $I_{\text{cascade }\gamma}\,(E>50$ GeV) contributes to the IGRB $_{\text{without unresolved sources}}(E>50$ GeV) about 10% for all spectral indices considered. Therefore in model I condition (1) is satisfied.

Neutrino fluxes are also generated during UHECR propagation. Another condition on CR models arises in accounting for neutrino fluxes; the cascade neutrino intensity $I_{\text{cascade }\nu}$ should be less than the intensity of astrophysics neutrino measured $I_{\nu \text{ measured}}$,

$$I_{\text{cascade }\nu} < I_{\nu \text{ measured}}. \qquad (6)$$

The neutrino intensity is measured at the neutrino observatory IceCube [31] and the PAO [32] in the energy range of about $(100 - 3 \cdot 10^6)$ GeV and $(2 \cdot 10^8 - 2 \cdot 10^{10})$ GeV, respectively. It appears that model I satisfies condition (6) both in the IceCube and in the PAO energy range, when the proton injection spectra have indices a > 0.5. Proton injection spectra with indices a = 0, 0.5 are formed either with a lower efficiency than in model I, or they are not realized. This is discussed in more detail in [8].

### 5. Results of model II

In model II in both cases of the source evolution, the calculated CR spectra are lower than the spectra measured by several orders of magnitude and differ greatly from them in shape. This is the same as in model I.

The model cascade gamma-ray intensity is the following:

$I_{\text{cascade }\gamma}$ ($E$>50 GeV, evolution of sources: BL Lac objects) = $1.002 \times 10^{-9}$ (cm$^{-2}$ s$^{-1}$ sr$^{-1}$), (7)

$I_{\text{cascade }\gamma}$ ($E$>50 GeV, evolution of sources: radio galaxies) = $1.641 \times 10^{-9}$ (cm$^{-2}$ s$^{-1}$ sr$^{-1}$). (8)

The difference in (7) and (8) results from various scenarios of source evolution.

Galactic diffuse emission models are used to derive the IGRB from the Fermi LAT data. Now we take into account the uncertainty of these models along with the measurement errors. We also account for the given error of $P$ of 14%, thus the contribution of unresolved gamma-ray sources is $P$= 72% (in contrast to $P$ = 86%, adopted in the estimate (5)). Then, we obtain the following band of IGRB$_{\text{without blazars}}$ ($E$ > 50 GeV):

$2.20 \cdot 10^{-10}$ (cm$^{-2}$ s$^{-1}$ sr$^{-1}$) $\leq$ IGRB$_{\text{without blazars}}$($E$ > 50 GeV)$\leq 5.40 \cdot 10^{-10}$ (cm$^{-2}$ s$^{-1}$ sr$^{-1}$). (9)

In this case the model intensity of the cascade gamma-ray emission is also higher than IGRB$_{\text{without blazars}}$($E$ > 50 GeV).

The calculated UHE CR flux is normalized to the flux measured at an energy of $10^{19.5}$ eV. However, the CR sources under consideration are 'exotic' ones so their contribution to the CR flux is definitely smaller, thus UHE CRs from these sources provide lower cascade emissions than (7), (8). In addition, UHE CRs emitted by other UHECR sources along with dark matter particle decays also contribute to IGRB. Now, we find the fraction R of the 'exotic' CR sources among BL Lac objects and radio galaxies using (1) and (5):

$R$ < 18% in comparison with BL Lac objects, (10)

$R$ < 11% in comparison with radio galaxies. (11)

To illustrate these estimates, we consider an example where the fraction of the 'exotic'

sources is 10% of the BL Lac objects and radio galaxies. Then the model intensity $I_{\text{cascade }\gamma}$ ($E > 50$ GeV) is:

$I\gamma$ ($E > 50$ GeV, evolution of sources: BL Lac objects) =

$$1.002 \times 10^{-10} \text{ (cm}^{-2}\text{ s}^{-1}\text{ sr}^{-1}) = 0.54 \cdot \text{IGRB}_{\text{without blazars}}, \quad (12)$$

$I\gamma$ ($E > 50$ GeV, evolution of sources: radio galaxies) =

$$1.641 \times 10^{-10} \text{ (cm}^{-2}\text{ s}^{-1}\text{ sr}^{-1}) = 0.89 \cdot \text{IGRB}_{\text{without blazars}}. \quad (13)$$

These values satisfy (1) and (5).

Values (12, 13) fall into the IGRB band (5). Moreover, there is room for the contribution of UHE CRs from the main sources and for a possible contribution of dark matter particle decays to the extragalactic diffuse gamma-ray emissions.

## 6. Discussion

Models I and II describe CR sources as contributing negligibly to the CR flux on Earth. Thus, another source provides the majority of UHE CRs, also generating electromagnetic cascades and providing the CR flux on Earth.

The minimal model of the majority of UHE CR sources was suggested in [6], describing both data on CRs and on extragalactic IGRB. The question is if model I is consistent with the model in [6].

In model I, cascade gamma rays contribute approximately 8-12% to the IGRB measured, thus leaving room for gamma rays initiated by the UHECR majority. Therefore, models I and [6] are in accordance.

In model II, we estimate the fraction of SMBHs surrounded by a superstrong magnetic field, to provide room for the contribution of UHE CRs from the main sources (and the possible contribution of dark matter particle decays to the extragalactic diffuse gamma-ray emissions).

In both models I and II, the proton spectra are normalized to satisfy the condition: the model particle flux does not exceed the measured one. However, it is unknown how much less can it be. Therefore, the gamma ray and neutrino fluxes obtained are the upper limits for cascade emission.

The central part of an AGN is surrounded by the gas–dust torus. We do not consider proton interactions with IR photons and gas in the torus. If particles interact in the torus, the production of secondary gamma rays and neutrinos increases [6].

Analyzing diffuse gamma ray intensity, the contribution from individual unresolved gamma ray sources should be extracted. At present the latter is determined by Fermi LAT, with the error of about 15% [30]. Gamma ray telescopes with better angular resolution than that of Fermi LAT are required to reduce this error. The parameters of current and planned gamma-telescopes are compared in [32]. The gamma ray telescope MAST has the best characteristics at energies above 20 GeV [32]; the gamma ray telescope GAMMA-400 [33] is also suitable for the investigation of CR sources.

Projects of neutrino observatories with improved parameters, compared with those of IceCube, are listed and discussed in [34] and references therein.

**7. Conclusion**

We consider that UHE CRs are accelerated in the vicinity of SMBHs, the CR injection spectra depend on processes of particle acceleration.

Following the results of [10], where CRs are accelerated in accretion discs and the maximal particle energy $E_{max}$ depends linearly on SMBH mass, we propose model I with CR injection spectra $\propto E^{-\alpha}$, where the spectral index a = 2.2, 1.8, 1, 0.5, 0; 0 corresponds to the equiprobable generation of particles at any UHE.

Following the ideas of [11], that SMBHs exist with a superstrong magnetic field and particles are accelerated in an induced electric field, we propose model II with a monoenergetic injection spectrum, the particle energy being $10^{21}$ eV.

In both models we analyze electromagnetic cascades arising when UHE CRs propagate in intergalactic space. It appears that possibly a subclass of extragalactic UHE CR sources exists, giving a very low particle flux on Earth, and which cannot be detected even with giant ground-based arrays. However, these UHE CRs produce noticeable fluxes of diffuse gamma rays and neutrinos in space. Therefore, the way to investigate this subclass of sources is to study extragalactic diffuse gamma ray and neutrino emissions.

In addition, we obtained an estimate of the fraction of possible 'exotic' SMBHs with superstrong magnetic fields.

Thus, in this paper we demonstrate that studying extragalactic electromagnetic cascades can be used as a tool to investigate UHE CR sources.

**Acknowledgment.** The author thanks O. Kalashev for discussion of the code TransportCR, M. Zelnikov for discussion of processes in SMBHs, and E. Bugaev for discussion of neutrino data obtained by PAO.